\documentclass[correspondence, draftcls, onecolumn, 12pt]{IEEEtran}
\usepackage[dvips]{graphicx}
\usepackage[fleqn]{amsmath}
\usepackage{subfigure} 
\usepackage{enumerate}

\begin{document}

\title{\huge A Study of Trade-off between Opportunistic Resource Allocation and Interference Alignment in Femtocell Scenarios}

\author{\IEEEauthorblockN{Namzilp Lertwiram \IEEEauthorrefmark{1}  \IEEEauthorrefmark{2},
Petar Popovski \IEEEauthorrefmark{1} and
Kei Sakaguchi \IEEEauthorrefmark{2}} \\
\IEEEauthorblockA{\IEEEauthorrefmark{1}Department of Electronic Systems, 
Aalborg University, Denmark\\
Email: \{namzilp, petarp\}@es.aau.dk}\\
\IEEEauthorblockA{\IEEEauthorrefmark{2}Tokyo Institute of Technology, Tokyo, Japan\\
Email: \{namzilp, sakaguchi\}@mobile.ee.titech.ac.jp}}

\maketitle

\begin{abstract}
One of the main problems in wireless heterogeneous networks is interference between macro- and femto-cells. Using Orthogonal Frequency-Division Multiple Access (OFDMA) to create multiple frequency orthogonal sub-channels, this interference can be completely avoided if each sub-channel is exclusively used by either macro- or a femto-cell. However, such an orthogonal allocation may be inefficient. We consider two alternative strategies for interference management,  opportunistic resource allocation (ORA) and interference alignment (IA). Both of them utilize the fading fluctuations across frequency channels in different ways. ORA allows the users to interfere, but selecting the channels where the interference is faded, while the desired signal has a good channel. IA uses precoding to create interference-free transmissions; however, such a precoding changes the diversity picture of the communication resources. In this letter we investigate the interactions and the trade-offs between these two strategies.
\end{abstract}

\section{Introduction}
A femtocell is a supplementary structure to a cellular network, implemented in areas where the signal from the base station (BS) cannot properly reach the users, especially ones in indoor areas. The emerging broadband wireless systems use  Orthogonal Frequency-Division Multiple Access (OFDMA) \cite{ofdma} to avoid interference among users. Due to the limited spectrum, allocating distinct sub-channels to all users in both macrocell and femtocells must be an inefficient way for interference management between macro-femtocells. 

In order to increase the spectrum efficiency, a common approach is to allow the femtocells to reuse the frequency band of the macrocell. A known interference issue when power control is applied to compensate pathloss of signal transmitted to a user at the cell edge is described as follows\cite{femto}. In downlink, this problem occurs when a macro user (M-UE) is located nearby a femto-user (F-UE) located at the cell edge of femtocell so that femto-BS (F-BS) has to raise the transmit power to reach this far F-UE, resulting in interference from the F-BS to the M-UE. 

Several strategies have been proposed to cope with interference problem in femtocell systems. This letter discusses two strategies that utilize fading fluctuations in frequency domain, i.e. Opportunistic Resource Allocation (ORA)\cite{ora1} and Interference Alignment (IA)\cite{IA1}-\cite{IA2}.
In ORA, with the variation of fading across different sub-channels, the system needs to find an appropriate sub-channel for a femtocell user, for which
this user has a high received power from his BS and less interference from
the macrocell transmission on the same sub-channel, such that total sum-rate is maximized.  On the other hand, IA utilizes fading fluctuations in frequency domain to generate pre-coding vectors which create interference-free channels (degrees of freedom). However, although sub-channels are assigned to users in the way that each user can gain the best signal from fading fluctuation, interference among users sharing the same resource is a key factor of system performance degradation. On the other hand, although IA utilizes frequency fading in order to create interference-free degrees of freedom, the fading fluctuations in this case are somehow averaged, thus suppressing the effect of favorable fading\cite{noise}. 

The different mechanisms behind ORA or IA, lead to the question which one is the best to be applied to utilize the limited spectrum resources in order to to maximize the system performance. This letter aims to investigate the trade-off between ORA and IA in the femtocell scenarios. Here, the available sub-channels are divided into two groups, i.e. ORA group and IA group, and the number of sub-channels allocated to ORA group is defined as \emph{trade-off number} between ORA and IA. By investigating this trade-off, we found that, in low SNR regime, the highest sum-rate can be achieved when most of sub-channels are allocated to perform ORA. On the other hand, when SNR is high, sub-channels are mostly allocated to perform IA in order to maximize the sum-rate. 

\begin{figure}
\begin{center}
\includegraphics[width=60mm]{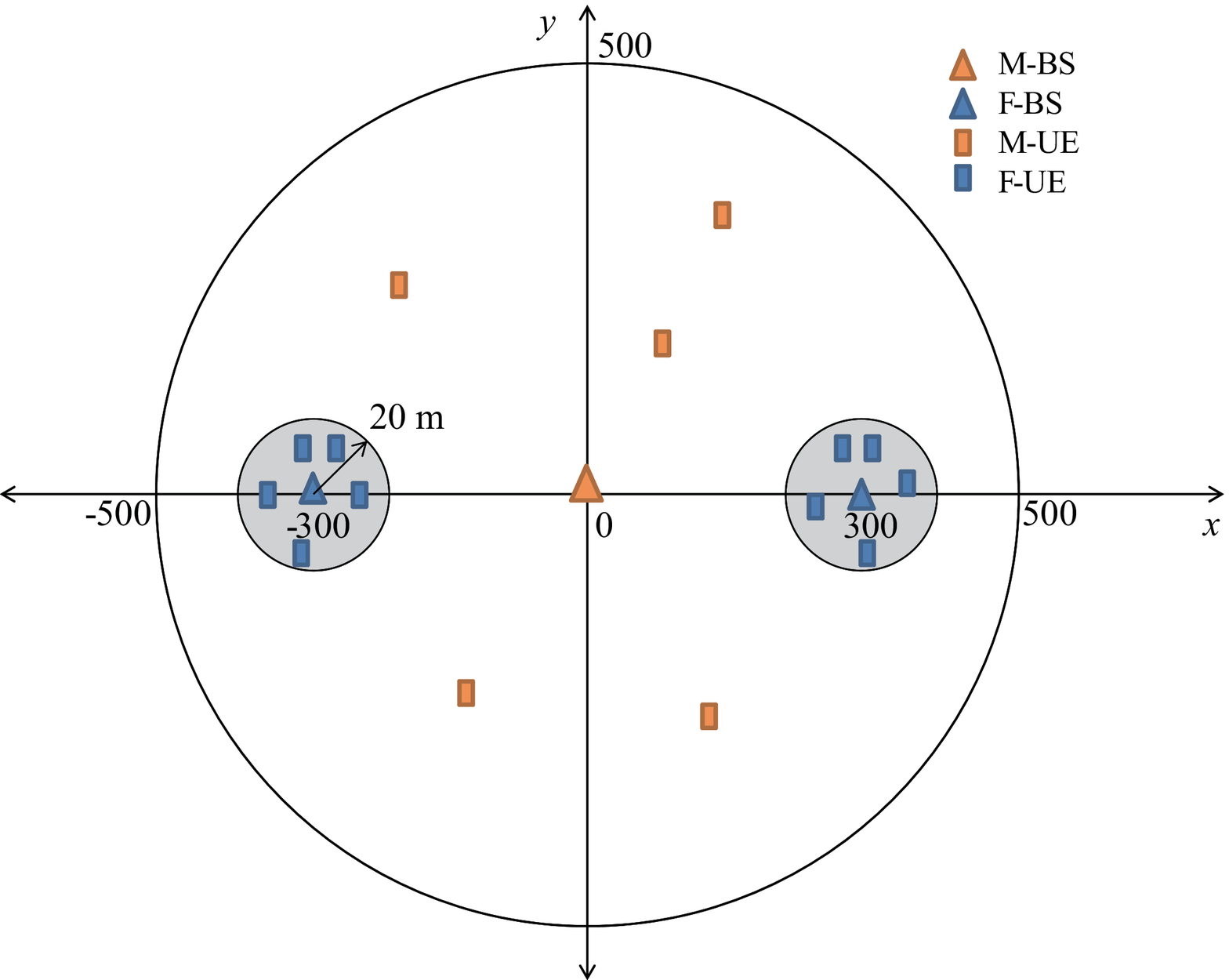}
\vspace{-3mm}
\caption{System model.}
\label{sysmodel}
\end{center}
\begin{center}
\includegraphics[width=60mm]{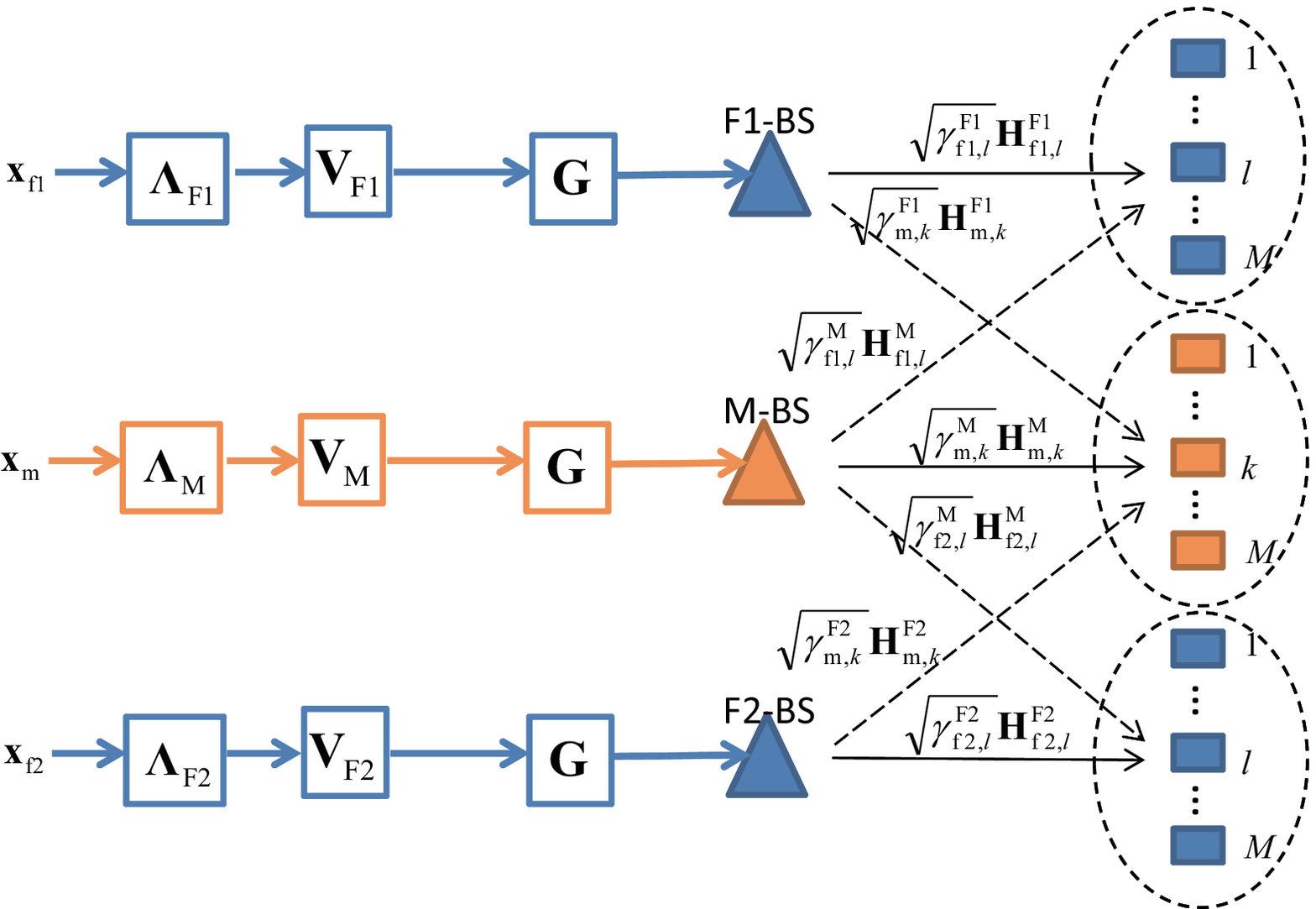}
\vspace{-3mm}
\caption{Interference alignment in a femtocell system.}
\label{fig_IA}
\vspace{-8mm}
\end{center}
\end{figure} 

\section{System Model}
We consider a cellular network consisting of two femtocells within a macrocell as shown in Fig.~\ref{sysmodel}. The frequency band in the macrocell is divided into sub-channels based on OFDMA and all sub-channels are reused by the femtocells. Only macro-femto interference is assumed, femto-femto interference is neglected\cite{PAmodel}. Here, each femtocell, as well as the macrocell, contains $L$ and $K$ users uniformly located in their corresponding BS coverage respectively, and all users share $N$ sub-channels. We assume that each sub-channel can only be allocated to a single user in each cell, but one user can have several allocated channels. When all users perform IA, the number of sub-channels must be assigned as $N \geq \max(L,K)+1$, as defined in \cite{Jafar}. In this letter, we assign $L = K = 5$ and $N = K + 1 = 6$ to facilitate our analysis.

\section{Channel Model}
Channel between a transmitter and a receiver is based on i. i. d. Rayleigh fading. We consider downlink transmissions only. $P_{\text{M}}$, $P_{\text{F1}}$ and $P_{\text{F2}}$ denote the transmission power of macro-BS (M-BS), femtocell 1-BS (F1-BS), and femtocell 2-BS (F2-BS) at each sub-channel, respectively. Receive power of macrouser (M-UE) $k$ from M-BS at sub-channel $n$ is $P^{\text{M}}_{\text{m},k,n} = P_{\text{M}} \lambda ^{\text{M}}_{\text{m},k} \zeta^{\text{M}}_{\text{m},k}  \mid h^{\text{M}}_{\text{m},k,n} \mid ^{2} $, where $h^{\text{M}}_{\text{m},k,n}$ is channel fading at sub-channel $n$, $\lambda^{\text{M}}_{\text{m},k}$ is the amplification factor for power control, and $\zeta^{\text{M}}_{\text{m},k} = K^{\text{M}}_{\text{m},k} (r^{\text{M}}_{\text{m},k})^{-\alpha}$ is path loss between M-BS to the user where $K^{\text{M}}_{\text{m},k} = \left( \frac{c}{4 \pi f_{c} d_{0}} \right)^{2}$ is unit-less path loss parameter varying with wavelength of RF carrier $c/f_{c}$ and reference distance $d_{0}$ of environment between BS to the user, $\alpha$ is path loss exponent for outdoor transmission, and $r^{\text{M}}_{\text{m},k}$ is the distance between M-BS to M-UE $k$\cite{PAmodel}. In case of indoor femtocell transmission, the received power of femtouser in femtocell 1 (F1-UE) $l$ from F1-BS at sub-channel $n$ is similarly determined as $P^{\text{F1}}_{\text{f1},l,n} = P_{\text{F1}} \lambda ^{\text{F1}}_{\text{f1},l} \zeta^{\text{F1}}_{\text{f1},l} \mid h^{\text{F1}}_{\text{f1},l,n} \mid ^{2} = P_{\text{F1}} \lambda ^{\text{F1}}_{\text{f1},l} K^{\text{F1}}_{\text{f1},l} (r^{\text{F1}}_{\text{f1},l})^{-\alpha} \mid h^{\text{F1}}_{\text{f1},l,n} \mid ^{2} $.

We assume perfect power control, adjusted to the propagation losses and ignoring the short-term fading effects. If sub-channel $n$ is allocated to M-UE $k$ in macrocell and F1-UE $l$ in femtocell1, $\lambda ^{\text{M}}_{\text{m},k} =  (r^{\text{M}}_{\text{m},k})^{\alpha}/K^{\text{M}}_{\text{m},k}$ and $\lambda ^{\text{F1}}_{\text{f1},l} =  (r^{\text{F1}}_{\text{f1},l})^{\alpha}/K^{\text{F1}}_{\text{f1},l}$. Therefore, the receive power at M-UE $k$ and F1-UE $l$ as desired signal become $P^{\text{M}}_{\text{m},k,n} = P_{\text{M}} \mid h^{\text{M}}_{\text{m},k,n} \mid ^{2}$ and $P^{\text{F1}}_{\text{f1},l,n} = P_{\text{F1}} \mid h^{\text{F1}}_{\text{f1},l,n} \mid ^{2}$, respectively. For indoor-outdoor and outdoor-indoor propagation, penetration loss from building walls $\delta $ is also considered. Hence, the receive power at M-UE $k$ from F1-BS and F1-UE $l$ from M-BS as interference signal become $P^{\text{F1}}_{\text{m},k,n} = P_{\text{F1}} (K^{\text{F1}}_{\text{m},k}/K^{\text{F1}}_{\text{f1},l}) (r^{\text{F1}}_{\text{m},k}/r^{\text{F1}}_{\text{f1},l})^{-\alpha} \delta \mid h^{\text{F1}}_{\text{m},k,n} \mid ^{2}$ and $P^{\text{M}}_{\text{f1},l,n} = P_{\text{M}} (K^{\text{M}}_{\text{f1},l}/K^{\text{M}}_{\text{m},k}) (r^{\text{M}}_{\text{f1},l}/r^{\text{M}}_{\text{m},k})^{-\alpha} \delta \mid h^{\text{M}}_{\text{f1},l,n} \mid ^{2}$, respectively. Since we consider perfect power control, the system performance determined in this paper is the upper-bound or maximum achievable performance of the network. 
  
\section{Channel Allocation Strategy for ORA-IA}
In order to study the trade-off between ORA and IA, the $N$ sub-channels are divided into two groups. The first group of $A$ channels is used for ORA and this number $A$ is briefly referred to as a \emph{trade-off number}. The second group of $N-A$ channels is used to perform IA. Basically, ORA searches for sub-channels in which there is least interference among the macro/femto transmissions, while IA is used to deal with strong interference among users in macrocell and femtocells. Therefore, channel assignment strategy for ORA and IA is in the way that $A$ sub-channels are allocated to users in ORA group prior to the rest $N-A$ sub-channels allocated to users in IA group, such that users in two groups can do their best with the different level of interference. In our analysis, we aim to find the optimum trade-off number which can make the network achieve the highest sum-rate with ORA-IA scheme.

From this point of view, it would be straightforward to think that the first $A$ sub-channels are allocated to $A$ users from each cell, and the rest $N-A = K-A+1$ sub-channels are allocated to the rest $K-A$ users from each cell. Here, the number of sub-channels and the number of users in each cell for both groups fit together. However, there are subtle problems with such an approach. When $A=N$, then we cannot consistently use the same ORA strategy, as there are $N = K+1$ and only $K$ users, such that there should be one user that gets allocated an additional sub-channel. 

Therefore, we need to define an ORA strategy that can be consistently applied for any $A \leq K+1$. This is done in the following way. First, we search for users, each from the macrocell and two femtocells, who can achieve the highest sum-rate at each channel. Then, we pick $A$ sub-channels which have the highest sum-rate among all $N$ sub-channels to be in ORA group. By this ORA strategy, we aim to determine the maximum achievable rate of users, so there might be some users assigned to more than one sub-channel. In addition, in the case that $A = K$, the number of sub-channels for IA group becomes $N-A = 1$ sub-channel which is not enough to perform IA, thus $A$ sub-channels are allocated to ORA group and IA is not performed in this case. 

The notations of sets used in allocation algorithm is defined as follows: $\mathcal{M}$, $\mathcal{F}_{1}$, $\mathcal{F}_{2}$ and  $\mathcal{N}$ denote the set of M-UEs, F1-UEs, F2-UEs and sub-channels respectively, and $k$, $l_{1}$, $l_{2}$ and $n$ denote the index of M-UE, F1-UE, F2-UE and sub-channel in sets $\mathcal{M}$, $\mathcal{F}_{1}$, $\mathcal{F}_{2}$ and $\mathcal{N}$ respectively. The following describes the procedure used for ORA. 
	\begin{enumerate}	
		\item  Initialization :   
		\begin{itemize}
			\item \footnotesize $\mathcal{M}^{\text{ORA}}$, $\mathcal{F}_{1}^{\text{ORA}}$, $\mathcal{F}_{2}^{\text{ORA}}$, $\mathcal{N}^{\text{ORA}} \leftarrow  \{ \}$;  \scriptsize\textsf{$\%\%$ Sets of ORA group}
			\item \footnotesize $\mathcal{M}^{\text{IA}}$, $\mathcal{F}_{1}^{\text{IA}}$, $\mathcal{F}_{2}^{\text{IA}}$, $\mathcal{N}^{\text{IA}} \leftarrow  \{ \}$; \scriptsize{\textsf{$\%\%$ Sets of IA group}}
		\end{itemize}\normalsize
		\item  Consider the rate of all users at each sub-channel without power control factor, e.g. M-UE $k$ at sub-channel $n$:\\  \footnotesize $C_{\text{m},k,n} = \log_{2}(1+ \frac{P_{\text{M}} \zeta^{\text{M}}_{\text{m},k}\mid h^{\text{M}}_{\text{m},k,n} \mid ^2 }{ P_{\text{F1}} \zeta^{\text{F1}}_{\text{m},k} \mid h^{\text{F1}}_{\text{m},k,n} \mid^2 + P_{\text{F2}} \zeta^{\text{F2}}_{\text{m},k} \mid h^{\text{F2}}_{\text{m},k,n} \mid^2 + \sigma^{2}})$.\\ \normalsize
		\item for $n = 1:N $ \footnotesize
\begin{eqnarray}
&&C_{\text{all},n,k,l_{1}, l_{2}} = C_{\text{m},k,n}+C_{\text{f1},l_{1},n}+C_{\text{f2},l_{2},n};  \nonumber  \\
&&C_{\text{max},n} = \max\limits_{\mathcal{M},  \mathcal{F}_{1},   \mathcal{F}_{2}}(C_{\text{all},n,k,l_{1}, l_{2} } ); \nonumber  \\
&&\{ k^{*[n]}, l^{*[n]}_{1}, l^{*[n]}_{2} \} \leftarrow \arg \max \limits_{\mathcal{M}, \mathcal{F}_{1}, \mathcal{F}_{2}}(C_{\text{all},n,k,l_{1}, l_{2} } ); \nonumber 
\end{eqnarray}
		\normalsize end for \scriptsize{\textsf{$\%\%$ Find the set of three users, each of which is from each cell, who make channel $n$ achieve the highest rate}} \normalsize \\			   
		\item \small $\mathcal{N}' \leftarrow \mathcal{N}$\normalsize \\
          for $a = 1:A $\footnotesize
			\begin{eqnarray*}
				 &&n^{*} = \arg \max\limits_{ \mathcal{N}'} (C_{\text{max},a}) \\
				 &&\mathcal{N}' \leftarrow \mathcal{N}' - \{ n^{*} \} \\
				 &&\mathcal{M}^{\text{ORA}}  \leftarrow k^{*[n^{*}]};  \mathcal{F}^{\text{ORA}}_{1}  \leftarrow l_{1}^{*[n^{*}]}; \\
			     &&\mathcal{F}^{\text{ORA}}_{2}  \leftarrow l_{2}^{*[n^{*}]}; \mathcal{N}^{\text{ORA}} \leftarrow n^{*}
			\end{eqnarray*}
			   \normalsize end for \scriptsize{\textsf{$\%\%$ Pick up the first $A$ sub-channels who achieve highest rate and allocate users and sub-channels in ORA group}} \normalsize \\
		\item  \footnotesize $\mathcal{M}^{\text{IA}} \leftarrow \mathcal{M} - \mathcal{M}^{\text{ORA}}; \mathcal{F}^{\text{IA}}_{1} \leftarrow  \mathcal{F}_{1} - \mathcal{F}^{\text{ORA}}_{1};$ \\ 
$\mathcal{F}^{\text{IA}}_{2} \leftarrow  \mathcal{F}_{2} - \mathcal{F}^{\text{ORA}}_{2}; \mathcal{N}^{\text{IA}}  \leftarrow \mathcal{N} - \mathcal{N}^{\text{ORA}}$ \\ \scriptsize{\textsf{$\%\%$ Let the rest users to be in IA group}} 
	\end{enumerate} \normalsize 
Note that, we perform the exhaustive search for sub-channels as described above in order to find the maximum sum-rate that the system can possibly achieve without the concern on the complexity in practical implementation. Also, the capacity of the selected users in each sub-channel is calculated again with power control factor. The remaining users in $\mathcal{M}^{\text{IA}} $, $\mathcal{F}^{\text{IA}}_{1} $ and $\mathcal{F}^{\text{IA}}_{2}$ are left to perform IA with the sub-channels in $\mathcal{N}^{\text{IA}}$. The details are described in the next section. 

\section{Interference Alignment Strategy for Femtocell Networks} 
In this section, we firstly review the strategy of IA in a femtocell network proposed in \cite{IA1}-\cite{IA2}. To facilitate the explanation, we consider the case that $J+1$ sub-channels are allocated to $J$ users in each cell to perform IA. At each BS (e.g. F1-BS), two precoders are applied to carry $J$ streams, $\mathbf{x}_{\text{f1}} = [x_{\text{f1,}1} ... x_{\text{f1},J}]^{T}$, each of which is intended for each user, over $J+1$ sub-channels, as shown in Fig. \ref{fig_IA}. The first precoder $\mathbf{G} \in \mathcal{C}^{J+1 \times J}$ is a fixed reference matrix which is a tall unitary matrix used for interference alignment. Here, we choose an orthogonal basis in $\mathcal{C}^{J+1 \times J}$ space for each column of $\mathbf{G}$, so that the columns of $\mathbf{G}$ are orthogonal and $\mathbf{G}^{H}\mathbf{G} = \mathbf{I}_{J \times J}$ which is a unitary matrix. The second precoder $\mathbf{V}_{\text{F1}} = [\mathbf{v}_{\text{f1},1} ... \mathbf{v}_{\text{f1},J} ] \in \mathcal{C}^{J \times J}$ is beamforming matrix which is used to decode the desired symbol at each user. For power adjustment of the received power at each user, the power control matrix $\mathbf{\Lambda}_{\text{F1}} = \text{diag}(\sqrt{\lambda ^{\text{F1}}_{\text{f1},1}},...,\sqrt{\lambda ^{\text{F1}}_{\text{f1},J}})$ is also applied before the precoders.

The received signal of F1-UE $l$ in the femtocell then becomes \small $\mathbf{y}_{\text{f1},l} = \sqrt{\zeta^{\text{F1}}_{\text{f1},l}} \mathbf{H}^{\text{F1}}_{\text{f1},l}\mathbf{G}\mathbf{V}_{\text{F1}}\mathbf{\Lambda}_{\text{F1}}\mathbf{x}_{\text{f1}} + \sqrt{\zeta^{\text{M}}_{\text{f1},l}} \mathbf{H}^{\text{M}}_{\text{f1},l}\mathbf{G}\mathbf{V}_{\text{M}}\mathbf{\Lambda}_{\text{M}}\mathbf{x}_{\text{M}} + \mathbf{z}_{\text{f1},l},$ \normalsize where \small $\mathbf{H}^{\text{F1}}_{\text{f1},l} = \text{diag}(h^{\text{F1}}_{\text{f1},l,1},...,h^{\text{F1}}_{\text{f1},l,J+1}) \in \mathcal{C}^{J+1 \times J+1} $ \normalsize denotes the direct channel from F1-BS to F1-UE $l$, \small $\mathbf{H}^{\text{M}}_{\text{f1},l} \in \mathcal{C}^{J+1 \times J+1}$ \normalsize denotes the cross-channel from M-BS, and \small $\mathbf{z}_{\text{f1},l} \in \mathcal{C}^{J+1 \times 1},\text{E} [ \mathbf{z}_{\text{f1},l}\mathbf{z}_{\text{f1},l}^{H} ]= \sigma^{2} \mathbf{I}_{J+1 \times J+1} $ \normalsize denotes the vector of additive white Gaussian noise at F1-UE $l$. Note that the received signal at each user in a femtocell 2 and macrocell can be considered in a similar manner.

Next, F1-UE $l$ estimates the interference channel from M-BS $\mathbf{H}^{\text{M}}_{\text{f1},l}\mathbf{G}$ by using a preamble, and then generates a null vector $\mathbf{u}_{\text{f1},l}\in \mathcal{C}^{J+1 \times 1}$, $\parallel \mathbf{u}_{\text{f1},l}\parallel ^{2} = 1$ such that $(\mathbf{u}_{\text{f1},l})^{H} \mathbf{H}^{\text{M}}_{\text{f1},l} \mathbf{G} = 0$. After applying this vector to the receive signal, outer cell interference is eliminated: $\tilde{y}_{\text{f1},l} = \sqrt{\zeta^{\text{F1}}_{\text{f1},l}}  (\mathbf{u}_{\text{f1},l})^{H}  \mathbf{H}^{\text{F1}}_{\text{f1},l} \mathbf{G} \mathbf{V}_{\text{F1}} \mathbf{\Lambda}_{\text{F1}} \mathbf{x}_{\text{f1}} + \tilde{z}_{\text{m},k}$, where $\tilde{z}_{\text{f1},l} =  (\mathbf{u}_{\text{f1},l})^{H} \mathbf{z}_{\text{f1},l}$. With this, there is still intra-cell interference left within the femtocell. In order to enable the F1-BS to cancel the interference among the femtocell users, each F1-UE feeds back the equivalent channel $(\mathbf{u}_{\text{f1},l})^{H}  \mathbf{H}^{\text{F1}}_{\text{f1},l} \mathbf{G}$ to F1-BS. Then, F1-BS calculates the beamforming matrix $\mathbf{V}_{\text{F1}}$ with the channel matrix $\mathbf{H}_{0}=[(\mathbf{u}_{\text{f1},1})^{H}\mathbf{H}^{\text{F1}}_{\text{f1},1} \mathbf{G} \hdots (\mathbf{u}_{\text{f1},J})^{H}\mathbf{H}^{\text{F1}}_{\text{f1},J} \mathbf{G} ]^{T}$. Note that in the case of macrocell, with the number of dimensions $J+1$, each M-UE is allowed to eliminate only one outer cell interference, so that M-BS performs IA to eliminate only the stronger interference between F1 BS and F2 BS and the rest is treated as noise.  

There is a freedom in choosing the transmit/receive vectors, such that we can optimize them in order to maximize the sum-rate achieved by IA. Therefore, we apply the MMSE-like algorithm proposed by \cite{IA1} as it provides the best sum-rate in any given SNR. With this scheme, the transmit/receive vectors of IA can be optimized to obtain a sum-rate higher than e.g. the simpler Zero-Forcing (ZF) based IA. At each BS (e.g. F1-BS), we first consider the covariance matrix of interference-plus-noise at user $l$ in the femtocell 1: \small $ \mathbf{\Phi }_{\text{f1},l} = \mathbf{I}\sigma^{2} + \frac{(M+1)P_{\text{M}}}{M} \zeta^{\text{M}}_{\text{f1},l} \mathbf{R}_{\text{f1},l},
$ \normalsize where \small $\mathbf{R}_{\text{f1},l} = \left( \mathbf{H}^{\text{M}}_{\text{f1},l} \mathbf{G} \mathbf{V}_{\text{M}} \mathbf{\Lambda}_{\text{M}} \mathbf{\Lambda}_{\text{M}}^{H} \mathbf{V}_{\text{M}}^{H}  \mathbf{G}^{H}  (\mathbf{H}^{\text{M}}_{\text{f1},l})^{H} \right)$.\normalsize

In uncordinated system, the precoding vector $\mathbf{V}_{\text{M}}$ is unknown to the users in the femtocell 1. Therefore, we use the expected value of the covariance matrix: \small $\mathbf{\bar {\Phi }}_{\text{f1},l} = \text{E} [ \mathbf{\Phi }_{\text{f1},l} ] = \mathbf{I}\sigma^{2} + \frac{(M+1)P_{\text{M}}}{M} \zeta^{\text{M}}_{\text{f1},l}\text{E}[\mathbf{R}_{\text{f1},l}]$\normalsize, where \small $\text{E}[\mathbf{R}_{\text{f1},l}] =  \left( \mathbf{H}^{\text{M}}_{\text{f1},l} \mathbf{G} \text{E} [ \mathbf{V}_{\text{M}} \mathbf{\Lambda}_{\text{M}} \mathbf{\Lambda}_{\text{M}}^{H}  \mathbf{V}_{\text{M}}^{H} ] \mathbf{G}^{H}  (\mathbf{H}^{\text{M}}_{\text{f1},l})^{H} \right)$\normalsize. Each entry in \small $\mathbf{V}_{\text{M}} = [\mathbf{v}_{\text{M},1} ... \mathbf{v}_{\text{M},J}]$ can be assumed as i.i.d. $\mathcal{C}\mathcal{N}(0,\frac{1}{K-A})$\normalsize, where \small $\text{E} [ \parallel \mathbf{v}_{\text{M},l} \parallel ^{2} ] = 1$ \normalsize is satisfied. Under this assumption, we obtain \small $\text{E} [ \mathbf{V}_{\text{M}} \mathbf{\Lambda}_{\text{M}} \mathbf{\Lambda}_{\text{M}}^{H}  \mathbf{V}_{\text{M}}^{H} ] = [\text{trace}(\mathbf{\Lambda}_{\text{M}} \mathbf{\Lambda}_{\text{M}}^{H}) /(K-A)] \mathbf{I}$\normalsize.

In our scenario, the number of users in sets $\mathcal{M}^{\text{IA}} $, $\mathcal{F}^{\text{IA}}_{1} $ and $\mathcal{F}^{\text{IA}}_{2}$ can be larger than $K-A$. Here, we do an opportunistic search again for $K-A$ users in each cell who achieve the highest performance  with MMSE-like algorithm IA algorithm as detailed below. (This procedure continues from step 5 in the previous section.)
\normalsize
\begin{enumerate} 
\setcounter{enumi}{5}
\item \normalsize for all \footnotesize $\mathcal{M}' \leftarrow \{ k'_{1},..,k'_{K-A} \in \mathcal{M}^{\text{IA}} \mid k'_{a} \neq k'_{b}, \forall a \neq b \}$ \\
 		\normalsize	for all \footnotesize $\mathcal{F}'_{1} \leftarrow \{ l'_{1,1},..,l'_{1,K-A} \in \mathcal{F}^{\text{IA}}_{1} \mid l'_{1,a} \neq l'_{1,b}, \forall a \neq b \}$ \\
 		\normalsize	for all \footnotesize $\mathcal{F}'_{2} \leftarrow \{ l'_{2,1},..,l'_{2,K-A} \in \mathcal{F}^{\text{IA}}_{2} \mid l'_{2,a} \neq l'_{2,b}, \forall a \neq b \}$ \\
\scriptsize \textsf{$\%\%$ For all possible subsets of users to perform IA with $K-A$ sub-channels} \\
\normalsize Each user initializes its receive vector, e.g. F1-UE $l$: \footnotesize $\mathbf{u}_{\text{f1},l}^{(0)} = \frac{\mathbf{\bar {\Phi }}_{\text{f1},l}^{-1}\mathbf{H}^{\text{F1}}_{\text{f1},l} \mathbf{G}\mathbf{v}_{\text{F1},l}^{(0)} }{\parallel  \mathbf{\bar {\Phi }}_{\text{f1},l}^{-1}\mathbf{H}^{\text{F1}}_{\text{f1},l} \mathbf{G}\mathbf{v}_{\text{F1},l}^{(0)} \parallel }$\normalsize, where $\mathbf{v}_{\text{F1},l}^{(0)}$ is set to be a maximum eigenvector of \footnotesize $\mathbf{G}^{H} (\mathbf{H}^{\text{F1}}_{\text{f1},l})^{H} \mathbf{\bar {\Phi }}_{\text{f1},l}^{-1} \mathbf{H}^{\text{F1}}_{\text{f1},l} \mathbf{G}$\normalsize. Then, each user feeds back the equivalent channel $(\mathbf{u}_{\text{f1},l}^{(0)})^{H}\mathbf{H}^{\text{F1}}_{\text{f1},l} \mathbf{G}$ to their own BSs and all BSs calculate ZF transmit vector \footnotesize $\mathbf{V}_{\text{F1}}^{(1)} = [\mathbf{v}_{\text{F1},1}^{(1)} ... \mathbf{v}_{\text{F1},J}^{(1)}] = \mathbf{H}_{0}^{H}(\mathbf{H}_{0}\mathbf{H}_{0})^{-1}\mathbf{D}$\normalsize, where \footnotesize $\mathbf{D} = \text{diag}(d_{1},..,d_{l},.,d_{J}), d_{l} = \frac{1}{\sqrt{\parallel (\mathbf{H}_{0}\mathbf{H}_{0}^{H})^{-1}\parallel_{l,l} }} $\normalsize.\\
\normalsize The rate of F1-UE $l$ can be calculated as: \footnotesize
\vspace{-0.5mm}
\begin{eqnarray*}
C_{\text{f1},l} = \log_{2} \left( 1+ \frac{1}{M}.\frac{(M+1)P_{\text{F1}}}{\sigma^{2}} \zeta^{\text{F1}}_{\text{f1},l}\parallel \mathbf{K}_{\text{f1},l} \parallel ^2 \right), 
\end{eqnarray*}
\normalsize where \footnotesize $\mathbf{K}_{\text{f1},l} = (\mathbf{u}_{\text{F1},l}^{(0)})^{H}  \mathbf{H}^{\text{F1}}_{\text{f1},l} \mathbf{G} \mathbf{v}_{\text{F1},l}^{(1)} \sqrt{\lambda^{\text{F1}}_{\text{f1},l}}$.\normalsize Finally, the sum-rate of users in $\mathcal{M}',\mathcal{F}'_{1},\mathcal{F}'_{2}$ performing IA with sub-channels in $\mathcal{N}^{\text{IA}}$ is calculated as \footnotesize
\begin{eqnarray}					
C_{\text{IA}}^{\mathcal{M}',\mathcal{F}'_{1},\mathcal{F}'_{2}} = \sum\limits_{ \mathcal{M}'}C_{\text{m},k} + \sum\limits_{\mathcal{F}'_{1}}C_{\text{f1},l_{1}} + \sum\limits_{\mathcal{F}'_{2}}C_{\text{f2},l_{2}} \label{sumIA}
\end{eqnarray}
\normalsize end for; end for; end for
\item \footnotesize $\{\mathcal{M}'^{*},\mathcal{F}'^{*}_{1},\mathcal{F}'^{*}_{2} \} \leftarrow \arg \max\limits_{\forall\mathcal{M}',\mathcal{F}'_{1},\mathcal{F}'_{2}}(C_{\text{IA}}^{\mathcal{M}',\mathcal{F}'_{1},\mathcal{F}'_{2}})$ \normalsize. \\ 
\scriptsize \textsf{$\%\%$ Pick up the set of users who make the network achieve the highest rate} \normalsize
\item For the set \footnotesize $\{\mathcal{M}'^{*},\mathcal{F}'^{*}_{1},\mathcal{F}'^{*}_{2} \}$\normalsize, the iteration of transmit-receive vector is performed, e.g. F1-UE $l$ updates the receive vector as $\mathbf{u}_{\text{f1},l}^{(i)} = \frac{\mathbf{\bar {\Phi }}_{\text{f1},l}^{-1}\mathbf{H}^{\text{F1}}_{\text{f1},l} \mathbf{G}\mathbf{v}_{\text{F1},l}^{(i)} }{\parallel  \mathbf{\bar {\Phi }}_{\text{f1},l}^{-1}\mathbf{H}^{\text{F1}}_{\text{f1},l} \mathbf{G}\mathbf{v}_{\text{F1},l}^{(i)} \parallel }$, where $i$ is iteration number.
Then, the equivalent channel is fed back again to BS to calculate the vector $\mathbf{v}_{\text{F1},l}^{(i+1)} $. Also, \footnotesize $C_{\text{IA}}^{\mathcal{M}''^{*},\mathcal{F}''^{*}_{1},\mathcal{F}''^{*}_{2} }$\normalsize is updated after the iteration.
\end{enumerate}

\section{Numerical Analysis}
\begin{figure}
\begin{center}
\includegraphics[width=85mm]{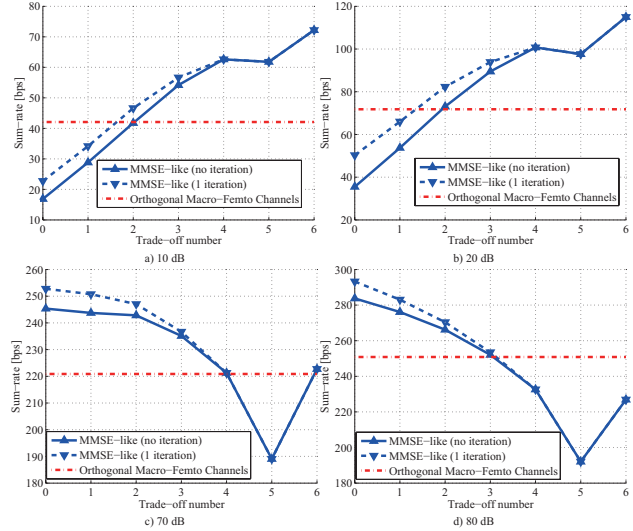}
\vspace{-3mm}
\caption{The sum-rate of the network at 10, 20, 70 and 80 dB.}
\label{suma}
\vspace{-8mm}
\end{center}
\end{figure} 

Computer simulation is used to analyze the sum-rate of the network with the following parameters: $P_{\text{M}} = 1$, $P_{\text{F1}} = P_{\text{F2}} = 1$, $\alpha=2$, $\delta=-10$ dB, $ d_{0} $ = 100 m (Outdoor), 5 m (Indoor), $f = 2$ GHz. Figure \ref{suma} shows the sum-rate of the network against trade-off number in the case when IA with MMSE-like without iteration and MMSE-like with iteration is applied, respectively. In these graphs, we also compare these sum-rate with the reference case when all 6 sub-channels are orthogonally divided into two groups, i.e. three sub-channels for M-UEs and the rest three sub-channels for F-UEs and each cell performs ORA to find its achievable sum-rate. We refer to this allocation as Orthogonal Macro-Femto Channels (OMFC). 

\begin{figure}
\begin{center}
\includegraphics[width=50mm]{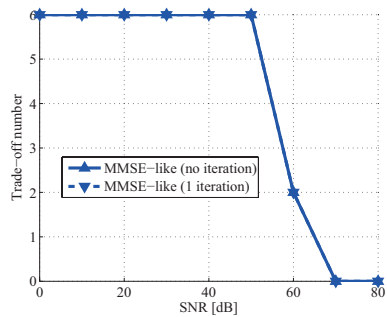}
\vspace{-3mm}
\caption{Trade-off number against SNR.}
\label{tradeoff}
\vspace{-6mm}
\end{center}
\end{figure}

In the case of low SNR, e.g. SNR = 10 and 20 dB, the system with ORA-IA can achieve its highest sum-rate when most sub-channels are allocated to perform ORA, as shown in Figs. \ref{suma}a and \ref{suma}b respectively. On the other hand, in the case of high SNR (SNR = 70 and 80 dB), the highest sum-rate can be obtained when most channels are allocated to perform IA as shown in Figs. \ref{suma}c and \ref{suma}d respectively. This is confirmed by observing the optimal trade-off number on Fig. \ref{tradeoff}. These results imply that the sum rate is maximized by applying an interference management strategy, i.e. switching between ORA and IA: opportunistically searching for the best channels with ORA can efficiently perform in low SNR regime, while interference alignment can show its performance advantage when SNR is higher. Note that the sum-rate when $A=5$ is always inferior to the case of pure ORA ($A=6$) since the network loose its chance to use the additional sub-channel which is reserved for IA but the system cannot perform IA in this case.

In addition, selecting the best between ORA and IA always outperforms OMFC. This suggests that reusing frequency resource by applying an appropriate interference management strategy such as ORA and IA at optimum trade-off number can improve the network performance rather than dedicating distinct resources for users in different cells to avoid interference.

By comparing the IA schemes with and without iterations, the figures show that the network performance can be increased when iteration is applied. However, the improvement with iteration is not significant in high SNR regime as shown in Fig. \ref{suma}d. This implies that the opportunistic search in IA should be followed by iterative update of the beamforming weights only in low SNR regime.

\section{Conclusion}
This letter discussed trade-off between ORA and IA in femtocell systems. The numerical result shows that the system tends to allocate most sub-channels to perform ORA and achieve the highest sum-rate in low SNR regime. On the other hand, the system tends to allocate more sub-channels to perform IA when SNR increases, while less sub-channels are allocated to ORA users. In our future work, we will investigate the general optimization approach for using OFDMA sub-channels with ORA and IA. 
 

\end{document}